\documentclass[12pt]{article}
\usepackage{amssymb}
\usepackage{amsmath}
\usepackage{amsfonts}

\usepackage{graphics}
\def\rlx{\relax\leavevmode}
\def\inbar{\vrule height1.5ex width.4pt depth0pt}
\def\IZ{\rlx\hbox{\small \sf Z\kern-.4em Z}}
\def\IR{\rlx\hbox{\rm I\kern-.18em R}}
\def\ID{\rlx\hbox{\rm I\kern-.18em D}}
\def\IC{\rlx\hbox{\,$\inbar\kern-.3em{\rm C}$}}
\def\IN{\rlx\hbox{\rm I\kern-.18em N}}
\def\IP{\rlx\hbox{\rm I\kern-.18em P}}
\def\one{\hbox{{1}\kern-.25em\hbox{l}}}

\def\beq{\begin{equation}}
\def\eeq{\end{equation}}
\def\bea{\begin{eqnarray}}
\def\eea{\end{eqnarray}}
\def\ber{\begin{array}}
\def\eer{\end{array}}

\begin{document}

\begin{titlepage}

October 2003 \hfill {UTAS-PHYS-03-07}\\

\mbox{}\hfill{\texttt{q-bio/}}
\vskip 1.6in
\begin{center}
{\Large {\bf $U(1)\! \times \! U(1) \times \! U(1)$ symmetry of the Kimura 
3ST model and phylogenetic branching processes }}
\end{center}

\normalsize
\vskip .4in

\begin{center}
J D Bashford$^{*}$, P D Jarvis$^{\dagger}$ and J G Sumner,  
\par \vskip .1in \noindent
{\it School of Mathematics and Physics, University of Tasmania}\\
{\it GPO Box 252-21, Hobart Tas 7001, Australia }\\[.3cm]
and M A Steel$^{\dagger}$,
\par \vskip .1in \noindent
{\it Biomathematics Research Centre, University of Canterbury,}\\
{\it Christchurch, New Zealand }\\
\end{center}
\par \vskip .3in \noindent

\vspace{1cm} An analysis of the Kimura 3ST model of DNA sequence
evolution is given on the basis
of its continuous Lie symmetries.  The rate matrix commutes with a
$U(1)\!  \times \!  U(1) \!  \times \!  U(1)$ phase subgroup of the
group $GL(4)$ of $4 \!  \times \!  4$ invertible complex matrices
acting on a linear space spanned by the 4 nucleic acid base letters. 
The diagonal `branching operator' representing speciation is defined,
and shown to intertwine the $U(1)\!  \times \!  U(1)\!  \times \! 
U(1)$ action.  Using the intertwining property, a general formula for
the probability density on the leaves of a binary tree under the
Kimura model is derived, which is shown to be equivalent to
established phylogenetic spectral transform methods. \\[.3cm]
\vfill
\hrule \mbox{} \\
{\footnotesize { $^{*}$ Australian Postdoctoral Fellow \\ 
$^{\dagger}$ Alexander von Humboldt Fellow}}
\end{titlepage}

The use of Markov models of stochastic change to taxonomic character
distributions is part of the standard armoury of techniques for
describing mutations and inferring ancestral relationships between
taxa.  For the simplest models, symmetries of the rate matrix under
discrete group actions (${\mathbb Z}_{2}$ for binary types, or
${\mathbb Z}_{2}\!  \times \!{\mathbb Z}_{2}$ for DNA or RNA bases in
molecular applications, for example) have been used to good effect in
simplifying phylogenetic analysis.  In particular, much attention has
been centred on properties of the frequently used Kimura 3ST
model\cite{Kimura3ST} which possesses such symmetry.

In this letter we describe an approach to the analysis of symmetries
using \textit{continous} transformation groups.  Rather than identify
the character types with elements of a (nonabelian or abelian)
discrete `colour' group which patterns the rate matrix for transitions
between types into orbit classes in the traditional way, we look at
linear transformations on the `character space' spanned by the
character types, and consider (complex, invertible) matrices which
\textit{commute} with the rate matrix.  As we shall show, this
approach, when implemented in the Kimura model, leads to an analysis
which is well adapted to the established Hadamard discrete Fourier
transform formalism\cite{HendyPenny,Erdos,HendyPennySteel,EvansSpeed},
but which importantly has potential generalisations going beyond the
colour groups.

Generically, let $\{ p_{a}(t), a = 1, \ldots, K \}$ be the 
probabilities that the system has trait
$a = 1,2,\ldots, K$ respectively. Introducing unit 
vectors  $e_{a}$, $a = 1,\ldots,K$ the state vector representing
the system 
\begin{equation}
        p(t) = p_{1}(t) e_{1} + p_{2}(t) e_{2} + \ldots + 
        p_{K}(t) e_{K}
        \label{eq:KstateProbVector}
\end{equation}
is subject to linear time evolution, 
\begin{equation}
	\label{eq:mastereq}
        \frac{d}{dt}{p}(t)  = \widehat{R} \cdot {p}(t),
\end{equation}
where the operator $\widehat{R}$ is a suitable $K \times K$ Markov rate
matrix. It is natural to decompose $\widehat{R}$ as
\begin{equation}
       \widehat{R} = \lambda (-\one + \widehat{T})
       \label{eq:RateMatrix}
\end{equation}
where the traceless part $\widehat{T}$ belongs by definition to the
Lie algebra $sl(K)$ (see below for the $K=4$ case). 
The usual (positive) rates for substitution
between different characters are thus the off-diagonal elements of
$\widehat{T}$.  A formal solution 
to (\ref{eq:mastereq}) for time-independent rates is
\begin{equation}
	\label{eq:oneqbitexpform}
	p(t) = e^{-\lambda t}\cdot e^{\lambda t\widehat{T}} \cdot p(0).
\end{equation}
The vector $p(t)$ in ${\mathbb C}^{K}$ is a probability density if
each $p_{a} \ge 0$ and $\sum_{a}p_{a} =1$.  Consistent with the time
dependence imposed by the master equation, given a starting density,
probability conservation is implemented by demanding that
$\widehat{R}$ is a unit column sum matrix.  Introducing the vector
$\Omega$ representing the sum of all the unit vectors in the
distinguished basis
\[
\Omega = e_{1}+e_{2}+ \ldots e_{K},
\]
probability conservation requires
that the dual $\Omega^{\perp}$ (the row vector with unit 
entries) is annihilated by $\widehat{R}$
regarded as an operator on the dual space, $\Omega^{\perp}
\cdot \widehat{R}=0$.  Equivalently, $\Omega^{\perp}$ is a
left \textit{unit} eigenvector of $\widehat{T}$.

In the Kimura model the characters $a$ are of course the standard 
nucleic acid base letters $A, G, U$ and $C$, and the rate matrix
is\footnote{
The rate parameters $\alpha, \beta, \gamma$ describe 
base transitions, and two classes of transversions, respectively.}
\begin{align}
	\left[\begin{array}{cccc}
	\widehat{R}_{AA} & \widehat{R}_{AG} & \widehat{R}_{AU} & \widehat{R}_{AC} \\
	\widehat{R}_{GA} & \widehat{R}_{GG} & \widehat{R}_{GU} & \widehat{R}_{GC} \\
	\widehat{R}_{UA} & \widehat{R}_{UG} & \widehat{R}_{UU} & \widehat{R}_{UC} \\
	\widehat{R}_{CA} & \widehat{R}_{CG} & \widehat{R}_{CU} & \widehat{R}_{CC} \end{array}\right] &= -(\alpha \!+\!\beta \!+\! 
	\gamma) \one +
	\left[ \begin{array}{cccc}
	0 & \alpha & \beta & \gamma \\
	\alpha & 0 & \gamma & \beta \\
	\beta & \gamma & 0 & \alpha \\
	\gamma & \beta & \alpha & 0 \end{array}\right] 
\end{align}
wherein the total change rate parameter in (\ref{eq:RateMatrix}) 
above is $\lambda = \alpha + \beta + \gamma$, and the traceless part 
of the rate operator can be written in the form
\begin{align}
	\widehat{T} &= \frac{\alpha}{\alpha + \beta + \gamma} \widehat{K}_{\alpha}
	+ \frac{\beta}{\alpha + \beta + \gamma} \widehat{K}_{\beta}+ 
	\frac{\gamma}{\alpha + \beta + \gamma} \widehat{K}_{\gamma}.
\end{align}
Remarkably the 3 Kimura matrices 
\begin{align}
	\widehat{K}_{\alpha} = \left[ \begin{array}{cccc}
	0 & 1 & 0 & 0 \\
	1 & 0 & 0 & 0\\
	0 & 0 & 0 & 1 \\
	0 & 0 & 1 & 0 \end{array}\right] , \quad
	\widehat{K}_{\beta} = \left[ \begin{array}{cccc}
	0 & 0 & 1 & 0 \\
	0 & 0 & 0 & 1\\
	1 & 0 & 0 & 0 \\
	0 & 1 & 0 & 0 \end{array}\right] , \quad
	\widehat{K}_{\gamma} = \left[ \begin{array}{cccc}
	0 & 0 & 0 & 1 \\
	0 & 0 & 1 & 0\\
	0 & 1 & 0 & 0 \\
	1 & 0 & 0 & 0 \end{array}\right] ,
\end{align}
provide a \textit{maximal set of commuting generators} for the Lie
algebra $sl(4)$ and thus can be chosen as the basis for a Cartan
subalgebra; equivalently there exists a transformation of the basis
spanned by $e_{A}, e_{G}, e_{U}$ and $e_{C}$ onto a new basis, in
which the Kimura generators are diagonal (with doubly degenerate
eigenvalues $\pm 1$ by the traceless property, and the fact that they
are square roots of $\one$).  This transformation is well known to be
generated by the Hadamard matrix ${\mathsf H}$, which sends
$\widehat{K}_{i}$ as matrices to ${\mathsf H} \widehat{K}_{i} {\mathsf
H}^{-1}$, $ i \in {\{} \alpha, \beta, \gamma {\}}$:
\begin{align}
	\label{eq:4by4similarity}
	\begin{array}{rr}
	\quad{\mathsf H}\!=\!\left[ \begin{array}{cccc}
	 1 &  1 &  1 &  1 \\
	 1 & -1 &  1 & -1 \\
	 1 &  1 & -1 & -1 \\
	 1 & -1 & -1 &  1 \end{array}\right] &, \quad
	{\mathsf H} \widehat{K}_{\alpha} {\mathsf H}^{-1} \!=\!\left[ \begin{array}{cccc}
	 1 & 0 & 0 & 0 \\
	0 & -1  & 0 & 0\\
	0 & 0 & 1 & 0 \\
	0 & 0 & 0 & -1 \end{array}\right], \nonumber \\ & \\  
	{\mathsf H} \widehat{K}_{\beta} {\mathsf H}^{-1} \!=\! \left[ \begin{array}{cccc}
	1 & 0 & 0 & 0 \\
	0 & 1  & 0 & 0\\
	0 & 0 & -1 & 0 \\
	0 & 0 & 0 & -1 \end{array}\right] &, \quad 
	{\mathsf H} \widehat{K}_{\gamma} {\mathsf H}^{-1} \!=\! \left[ \begin{array}{cccc}
	1 & 0 & 0 & 0 \\
	0 & -1  & 0 & 0\\
	0 & 0 & -1 & 0 \\
	0 & 0 & 0 & 1 \end{array}\right].\end{array}
\end{align}
Note finally that $\mathsf H$ can be decomposed as a tensor product of two-dimensional forms
\begin{align}
    	\mathsf H = {\mathsf h} \otimes {\mathsf h}, 
		\qquad {\mathsf h} = \left[ \begin{array}{cc} 1 & 1 \\  
		1 & - 1 \end{array} \right].
\end{align}
		
Thus far, we have recovered the standard analysis, with the 
emergence of the Hadamard transformation as the key to resolving the Kimura 
model. For (multi)-taxon probability 
densities \textit{evolving independently}, the time evolution after 
time $t$ is by extension of (\ref{eq:oneqbitexpform})
\begin{equation}
\label{eq:tensorproductform}
	P(t) =  \displaystyle{ e^{-\lambda t} \cdot e^{\lambda t 
	\widehat{\mathbb T}} } \cdot  P(0),
\end{equation}
where $P$ is a tensor of rank $\ge 2$ carrying the
probability density on a sample space spanned by the appropriate Cartesian
product of character sets, and $\widehat{\mathbb T} := \widehat{T}  
\otimes   \one   \otimes  \ldots \one + \one   \otimes 
\widehat{T}\ldots + \ldots $ the corresponding off-diagonal rate operator lifted to
the tensor product space.  Clearly, the higher rank Hadamard operator
${\mathsf H} \!  \otimes \!  {\mathsf H}\!  \otimes \ldots$ again implements
the correct diagonalisation in this case.  The work of
\cite{HendyPenny,Erdos,HendyPennySteel,EvansSpeed} using discrete Fourier analysis on
trees establishes that, remarkably, the Hadamard transform technique
\textit{still} applies, even when the multi-taxon system has evolved via a
phylogenetic tree.

In order to pursue the alternative analysis via Lie symmetries, we
exploit the observation that the Kimura operators $\widehat{K}_{i}$,
$i \in {\{} \alpha, \beta, \gamma {\}}$ provide a Cartan subalgebra
for transformations belonging to the Lie algebra $sl(4)$ of the group
$SL(4)$ of (complex) matrices\footnote{$SL(4) \simeq GL(4)/{\mathbb
C^{\times}}$ where invertible matrices are factored by the
multiplicative group of complex numbers corresponding to their
(nonzero) determinants.}.  Clearly the Hadamard basis vectors $h_{a}
:= {\mathsf H} \cdot e_{a}$ are simultaneous eigenvectors of the
Kimura generators.  In a general representation of $SL(4)$, the
eigenvalues of the Kimura generators simply correspond to the weight
decomposition with respect to the Cartan subalgebra.  Aside from the 
overall scaling by $ e^{-\lambda t}$, the Markov transition
operator $ e^{-\lambda t}\cdot e^{\lambda t\widehat{\mathbb T}}$
appends an exponential time dependence given by the sum of these
weights, multiplied by the Kimura `charge' parameters $\alpha, \beta,
\gamma$.  Thus, in principle, provided the Markov model respects the
symmetry, its spectral properties, and hence the time development of a
multi-taxon density, can be deduced from an appropriate weight
decomposition of the corresponding tensor representation of $SL(4)$.

To confirm that the analysis does indeed carry through in the presence of
phylogenetic trees, we now turn to the description of the branching
process itself.  The usual formalism of stochastic models of base
substitution\cite{Rodriguez} can conveniently be encapsulated via a
linear operator $\delta$, which changes the state vector representing
a single taxon, to that representing independent progeny after
branching\footnote{A dynamical, many-body formulation of phylogenetic
branching processes has been presented in \cite{JarvisBashford2000}.}.  In the
nucleic acid basis we have
\begin{align}
	\label{eq:deltadefn}
\delta \cdot e_{A} = e_{A}\otimes e_{A}, \quad & \quad \delta \cdot 
e_{G} = e_{G}\otimes e_{G}, \nonumber \\
\delta \cdot e_{U} = e_{U}\otimes e_{U}, \quad & \quad \delta \cdot 
e_{C} = e_{C}\otimes e_{C},
\end{align}
so that, when applied to a vector $p$ representing the density on 
bases for one taxon, we have 
\begin{align}
	\label{eq:deltaaction}
	p = & \, p_{A} e_{A} + p_{G} e_{G}
	+ p_{U} e_{U} + p_{C} e_{C} \nonumber \\
	\rightarrow \, 
	\delta \cdot p = & \, p_{A} e_{A}\otimes e_{A} + p_{G} e_{G}
	\otimes e_{G}+ p_{U} e_{U}\otimes e_{U} + p_{C} e_{C}\otimes e_{C},
	\end{align}
after which evolution proceeds for the model on two taxa (with all 
operations lifted to the tensor product space carrying the Cartesian 
square of the character set, as described by 
(\ref{eq:tensorproductform}) above).

A stochastic model may be said to possess a \textit{symmetry} under a 
continuous transformation group $G$ \textit{if the rate matrix commutes with its 
generators}, and hence intertwines the group 
action. Formally if the action is $p(t) \rightarrow p'(t) \equiv g\cdot p(t)$ 
then the master equation (\ref{eq:mastereq})
retains its form, for all $g \in G$ and for arbitrary $p(t)$, as
\[
\frac{dp'(t)}{dt} =  \widehat{R} \cdot p'(t)
\]
iff $g\widehat{R} = \widehat{R}g$, or ${[}\widehat{R}, 
\widehat{K}{]} =0 $ with $\widehat{K}$ a \textit{generator} of the group $G$ 
($g \sim e^{\widehat{K}}$). Similarly a branching operator $\delta$ admits 
a symmetry under such transformations if it intertwines the action of 
$G$ on the character space of a single taxon, with some action on the 
tensor product space:
\begin{align}
	\label{eq:coproduct}
	\delta \, {\scriptstyle \circ}\, g &=  \widetilde{g} \, {\scriptstyle 
	\circ} \, \delta.
\end{align}

Such symmetry considerations lead to useful ways of analysing the tree
structure of general phylogenetic branching processes, which we hope
to take up in a separate work.  Here we examine
the implications for the Kimura model as a first example.  It is clear
from the above remarks that the rate matrix admits a $GL(1) \times
GL(1) \times GL(1)$ $\simeq$ ${\mathbb C}^{\times} \times {\mathbb
C}^{\times} \times {\mathbb C}^{\times}$ group of symmetry
tansformations.  For present purposes it is sufficient to take the
corresponding unitary phase subgroup, $U(1)  \times  U(1)  
\times U(1)$.  Turning to the diagonal branching operator
(\ref{eq:deltadefn}), it is obvious that the Kimura generators in the
distinguished basis simply act as permutations of the
basic unit vectors $e_{A}, e_{G}, e_{U}, e_{C}$ and hence themselves
have a diagonal intertwining property\footnote{In the case of abelian
algebras, a `group-like' coproduct $\widehat{K} \rightarrow
\widehat{K}\!\otimes \!  \widehat{K}$ gives a coassociative coalgebra
structure, and the tensor product spaces carry a consistent comodule
action.}:
\begin{equation}
	\label{eq:groupcoproduct}
	\delta \, {\scriptstyle \circ} \, \widehat{K}_{i} = 
	\widehat{K}_{i} \otimes \widehat{K}_{i} \, {\scriptstyle \circ} \, \delta, \quad i \in 
	{\{}\alpha, \beta, \gamma {\}}.
\end{equation}
Thus we conclude that the Kimura model has $U(1) \!  \times \!  U(1)
\!  \times \!  U(1)$ symmetry, \textit{both} in the sense of commuting
with the rate matrix, \textit{and} in the intertwining property for
the branching operator.

With the above preliminaries we sketch briefly the the way in which
the above algebraic structure can be applied to an analysis of the
Kimura model for phylogenetic trees, which is consistent with the
Fourier transform methods.  Fixing a rooted tree on $L$ leaves,
the full time evolution from the initial root density to the leaf
density can be represented abstractly as a product of strings of
terms of the form
\begin{equation}
\label{eq:stringproduct}
\ldots(M_{1}'\otimes M_{2}'\otimes \ldots \otimes M_{r+1}')\cdot (\one \otimes \one 
\otimes \ldots \delta \otimes \ldots \otimes \one)\cdot
(M_{1}\otimes M_{2}\otimes \ldots \otimes M_{r})\ldots
\end{equation}
where it is implied that, for the time slices of the tree under consideration,
with $r$ taxa evolving, a branching event\footnote{
See also \cite{JarvisBashford2000}.} took place on a particular edge
leading to $r+1$ taxa evolving, the $M$ being simply the appropriate
Markov transition matrices $e^{\Delta t \widehat{R}}$. The intertwining property 
(\ref{eq:groupcoproduct}) can now be used to pull all the $\delta$ 
operators back to the root node, so that the final expression for the 
leaf density is of the form of products of exponentials of tensor products 
of Kimura operators, acting on the fully branched state\footnote{
The operator $\delta$ is coassociative, $(\one \otimes 
\delta){\scriptstyle \circ}\delta = (\delta \otimes \one 
){\scriptstyle \circ}\delta \equiv  \delta^{(2)}$.}
\begin{align}
\label{eq:GHZstate}
\delta^{(L-1)}p(0) & = p_{A}(0) e_{A}\otimes e_{A} \ldots \otimes e_{A} + 
p_{G}(0) e_{G}\otimes e_{G} \ldots \otimes e_{G} + 
\nonumber \\
& p_{U}(0) e_{U}\otimes 
e_{U} \ldots \otimes e_{U}
+ p_{C}(0) e_{C}\otimes e_{C} \ldots \otimes e_{C}.
\end{align}
Working in the Hadamard basis allows the exponentials to be 
diagonalised in terms of the weights of the tensor product states 
under the induced $U(1)\! \times \! U(1) \! \times \! U(1)$ action. The 
combinatorics of the tree is of course encoded, in that the change on 
each edge explicit in (\ref{eq:stringproduct}) is inherited by the 
differing total weights of each factor, and hence different 
exponential time dependence, in the $L$ edges emanating from  
(\ref{eq:GHZstate}) above.

As an example we specialise to the binary character case (the
symmetric two colour model \cite{Farris,Cavender}).  Suppose the character
set is ${\{} Y, R {\}}$ for definiteness.  The analogue of the Kimura
operator is $\widehat{\mbox{k}}$, and there is only one rate parameter
$\alpha$ with $\widehat{R} = \alpha(-\one + \widehat{\mbox{k}})$.  The
analogue of (\ref{eq:4by4similarity}) is
\begin{align}
	\label{eq:2by2similarity}
	\widehat{\mbox{k}} = \left[ \begin{array}{cc} 0 & 1 \\ 1 & 0 \end{array}\right], 
	\quad & {\mathsf h} = \left[ \begin{array}{cc}  1 &  1 \\ 
		1 & - 1 \end{array} \right], \quad {\mathsf h} \widehat{\mbox{k}} 
		{\mathsf h}^{-1} = \left[\begin{array}{cc} 1 & 0 \\ 0 & -1 \end{array}\right].
	\end{align}
Consider the descending rooted 4-leaf tree $(1(2(34)))$. 
Labelling the non-leaf edges $5,6$ in order 
of ascending level away from the leaves, define the total edge change parameters (including 
time intervals) as 
\[  \alpha_{e} := {\Delta 
t}_{e}\alpha, \quad e \in {\{} 1,2, \ldots, 6 {\}}
\]
(effectively allowing the $\alpha$ parameter in the rate matrix to be 
edge-dependent), and also the leaf operators
\begin{align}
\widehat{\mbox{k}}_{1}= \widehat{\mbox{k}}\otimes \one  \otimes \one  \otimes \one , 
\quad & \quad
\widehat{\mbox{k}}_{2}= \one \otimes \widehat{\mbox{k}} \otimes \one \otimes \one, 
\nonumber \\
\widehat{\mbox{k}}_{3}=  \one  \otimes \one \otimes \widehat{\mbox{k}}  \otimes \one , 
\quad & \quad
\widehat{\mbox{k}}_{4}= \one \otimes \one \otimes \one \otimes 
\widehat{\mbox{k}}. \nonumber
\end{align}
Applying (\ref{eq:groupcoproduct}), (\ref{eq:stringproduct}), we have
for the leaf density
\begin{align}
	\label{eq:treeopex}
	 P_{leaf} =  e^{-\sum_{e}\alpha_{e}} & \cdot 
	 e^{  \alpha_{1}\widehat{\mbox{k}}_{1} + \alpha_{2}\widehat{\mbox{k}}_{2} + 
	 \alpha_{3}\widehat{\mbox{k}}_{3} + \alpha_{4}\widehat{\mbox{k}}_{4} + 
	 \alpha_{5} \widehat{\mbox{k}}_{5}  + \alpha_{6} \widehat{\mbox{k}}_{6} }
	 \cdot \delta^{3} p(0) , \nonumber \\
	 \mbox{where also} \quad 
	 \widehat{\mbox{k}}_{5} = \, \one \otimes \one \otimes \widehat{\mbox{k}} 
	 \otimes \widehat{\mbox{k}}, \, 
	 & \qquad \widehat{\mbox{k}}_{6} = 
	 \one  \otimes \widehat{\mbox{k}} \otimes \widehat{\mbox{k}} \otimes \widehat{\mbox{k}}.   
     \end{align}
The composite operator in (\ref{eq:treeopex}) acts in the Hadamard basis to give a 
signed sum of edge parameters, with the signs determined by products 
of $\widehat{\mbox{k}}$-weights, eigenvalues of the various leaf 
operators acting on $\delta^{3}\! p(0) =  
p_{Y}(0) e_{Y}\otimes e_{Y}\otimes e_{Y}\otimes e_{Y} + p_{R}(0) 
e_{R}\otimes e_{R}\otimes e_{R}\otimes e_{R}$ expanded via the inverse Hadamard 
transform (see (\ref{eq:2by2similarity})), 
\begin{align}
	\label{eq:etohoverlap}
	e_{Y} = \frac{1}{2} (h_{+}+ h_{-}), \quad & \quad  e_{R} = \frac{1}{2} 
	(h_{+}- h_{-}).
\end{align}
Multiplying through by the overall prefactor, the \textit{positively} signed edge 
parameters cancel in the exponent. For example the 
coefficient of $h_{+} \otimes h_{-} \otimes h_{+} \otimes h_{-}$ in 
the expansion of (\ref{eq:treeopex}) becomes
\[ 
P_{+-+-} = \textstyle{{(\frac{1}{2})}^{4}}\cdot e^{-2 (\alpha_{2}+ 
\alpha_{4}+ 
\alpha_{5})}p_{Y}(0) + \textstyle{ 
{(\frac{1}{2})}^{2}\cdot  
{(-\frac{1}{2})}^{2}} \cdot e^{-2 (\alpha_{2}+ \alpha_{4}+ 
\alpha_{5})}p_{R}(0).
\]

As explained above, the use of (\ref{eq:groupcoproduct}),
(\ref{eq:stringproduct}) in generalising (\ref{eq:treeopex}) to an
arbitrary tree ${\mathcal T}$ amounts to considering how the symmetry
group on the linear space spanned by the evolving probability density
of a single system, extends after branching to transformations acting
on the $L$-fold tensor product (in the Kimura 3ST model, the symmetry
group is $U(1)\times U(1)\times U(1)$, and in the symmetric two-colour
model just $U(1)$).  Taking the binary case for simplicity, the general
form of (\ref{eq:treeopex}) reads (\textit{cf}
(\ref{eq:tensorproductform}) and (\ref{eq:oneqbitexpform}))
\begin{align}
\label{eq:treeform}
P_{leaf} = & e^{-\sum_{e}{\alpha_{e}}} \cdot e^{\widehat{\mbox{k}}_{\mathcal T}} \cdot 
\delta^{(L-1)}\!p(0).
\end{align}
The operator 
$\widehat{\mbox{k}}_{\mathcal T}$ is the induced generator of $U(1)$ after pulling 
back through the branching nodes of the tree. We define (following the 
above example)
\[
\widehat{\mbox{k}}_{(e)} = \prod_{\ell \in {\mathcal T}_{e}} \widehat{\mbox{k}}_{\ell}
\]
for each edge $e$ to be the product, over all leaves in the subtree ${\mathcal T}_{e}$ 
determined by $e$, of the leaf operators 
\[
\widehat{\mbox{k}}_{\ell} = 1  \otimes 1 \otimes \ldots \widehat{\mbox{k}} \otimes 
1 \otimes \ldots \otimes 1
\]
($\widehat{\mbox{k}}$ acting on the $\ell$'th place in the $L$-fold tensor 
product -- obviously if $\ell$ is a leaf edge, 
$\widehat{\mbox{k}}_{(\ell)}\equiv \widehat{\mbox{k}}_{\ell}$). Then
\[
\widehat{\mbox{k}}_{\mathcal T} = \sum_{e} \alpha_{e} \widehat{\mbox{k}}_{(e)}.
\]
Finally, $\delta^{L-1}p(0)$ is the maximally branched state (as would 
derive from a multifurcating branching). Note however, that this 
decomposition does not imply that the leaf density is equivalent 
to independent stochastic evolution from this initial branched 
state -- the operator $\widehat{\mbox{k}}_{\mathcal T}$ is not of separable 
form.

While (\ref{eq:treeform}) is independent of basis, it is obviously
beneficial to analyse the components of each side in terms of the
(tensor products of) Hadamard vectors (eigenstates of the
$\widehat{\mbox{k}}$ operator), as both the separable and
non-separable parts of the tree operator $\widehat{\mbox{k}}_{\mathcal
T}$ are diagonal in this basis.  Briefly the algorithm for determining
the weight attributed to a term of $P_{leaf}$ in the Hadamard basis
can be described as follows (for a formal analysis see
\cite{JarvisSumnerBashford2003}).  Take an arbitrary binary tree, and fix
a tree `split', to be associated with the coefficient, in the expansion
of $P_{leaf}$, of the basis element consisting of the $L$-fold tensor
product of $-$ Hadamard vectors on a chosen subset of distinguished
leaves, with $+$ Hadamard vectors at the remaining non-distinguished leaf
positions.  On the graph of the tree assign $-$ signs to the
distinguished leaf edges, and $+$ signs to the remainder, and
propagate signs to the remaining edges multiplicatively (\textit{eg}
adjacent siblings with $-$ signs will generate a $+$ sign on
their ancestral edge).  The corresponding signed sum of edge
parameters $\alpha_{e}$ is precisely the exponent generated by the action
of $\widehat{\mbox{k}}_{\mathcal T}$ on this basis element.  After the
overall $e^{-\sum_{e} {\alpha_{e}}}$ prefactor is multiplied through,
\textit{only} the negatively signed edge terms are present in the
exponent (with coefficient -2).  Finally the numerical factors
accompanying the terms proportional to $p_{Y}(0)$ and $p_{R}(0)$ can
easily be read off from (\ref{eq:etohoverlap}).

It is clear that the above presentation is equivalent to the standard
discrete Fourier analysis on trees techniques involving the Hadamard
transform\cite{HendyPenny,Erdos,HendyPennySteel,EvansSpeed}.  Specifically, the
surviving edge parameters which provide the argument of the
exponential are nothing but the nonintersecting path edge sums for a
given leaf split, as emerges from the Hadamard transform in edge
space.  The standard ${\mathbb Z}_{2}\times {\mathbb Z}_{2}$ colour
symmetry is of course inherent in the Hadamard matrix, which is also
mandatory for the simultaneous diagonalisation of the Kimura
generators.  However, from the viewpoint of Lie symmetries, the latter
determine 3 (infinite) continuous symmetry groups, rather than being
identified with the 3 non-unit elements of a discrete group.  Crucial
for our derivation is the coproduct property (\ref{eq:coproduct}), and
the fact that the combinatorics of the tree determines the final
action of the symmetry group on the $L$-fold tensor product carrying
the leaf probability density.

In this letter we have provided a framework for the analysis of 
phylogenetic branching models on the basis of continuous transformation 
symmetries of the rate matrix and the branching operator. The 
formalism can be applied to the Kimura 3ST (and also the 2P) 
model, as well as the symmetric binary character model 
\cite{Farris,Cavender} and it reproduces the standard spectral 
transform analysis. Importantly, it can potentially be applied to any 
model where the off-diagonal rates can be associated with an 
abelian subalgebra of $SL(K)$ whose generators have the form of 
permutation matrices (so that the intertwining property holds). We 
defer a formal presentation of such generalisations, and of the role of Lie symmetries and 
representation theory in branching models to a separate 
paper\cite{JarvisSumnerBashford2003}.

\subsubsection*{Acknowledgements}
PDJ and JGS thank the Department of Physics and Astronomy, and also
the Biomathematics Research Centre, University of Canterbury, Christchurch, New 
Zealand, for hosting a visit during which this work was initiated.
This research was supported by the Australian Research Council  
grant DP0344996.

 \end{document}